\documentclass[12pt, preprint]{aastex} 

\usepackage{graphicx}

\shorttitle{A Young Eclipsing Binary and its Luminous Neighbors in Sh 2-252E}
\shortauthors{Lester et al.}

\begin{document}
\title{A Young Eclipsing Binary and its Luminous Neighbors in the Embedded Star Cluster Sh 2-252E}
\author{Kathryn V. Lester,  Douglas R. Gies, Zhao Guo}
\affil{Center for High Angular Resolution Astronomy and Department of Physics \& Astronomy, \\ Georgia State University, P.O. Box 5060, Atlanta, GA 30302-5060, USA}
\email{lester@chara.gsu.edu, gies@chara.gsu.edu, guo@chara.gsu.edu}

\begin{abstract}

We present a photometric and light curve analysis of an eccentric eclipsing binary in the K2 Campaign 0 field that resides in Sh~2-252E, a young star cluster embedded in an H~II region. We describe a spectroscopic investigation of the three brightest stars in the crowded aperture to identify which is the binary system. We find that none of these stars are components of the eclipsing binary system, which must be one of the fainter nearby stars. These bright cluster members all have remarkable spectra: Sh~2-252a (EPIC~202062176) is a B0.5~V star with razor sharp absorption lines, Sh~2-252b is a Herbig A0 star with disk-like emission lines, and Sh~2-252c is a pre-main sequence star with very red color.

\end{abstract}
\keywords{binaries: eclipsing, stars: massive, open clusters and associations: individual (Sh 2-252E), stars: pre-main sequence, stars: variables: T Tauri, Herbig Ae/Be}
\section{Introduction}
Time series photometry from the Kepler K2 mission \citep{howell14} has led to the discovery of a number of interesting eclipsing binary systems. A particular example and the subject of this paper is an eccentric orbit binary associated with EPIC~202062176 \citep{armstrong15, lacourse15}. This star sits at the center of a young cluster of stars embedded in the H~II region Sh~2-252E, which in turn is one of six young clusters associated with the large, diffuse H~II region, NGC~2174 \citep[Sh 2-252,][]{sharpless59, bonatto11}. All these clusters have an estimated age of 5~Myr or less, consistent with the age of the primary ionizing source, the O6.5~V star HD~42088 \citep{glz71, felli77, jose12}.

\begin{figure}[tb!]  
\centering
\epsscale{1.0}
\plotone{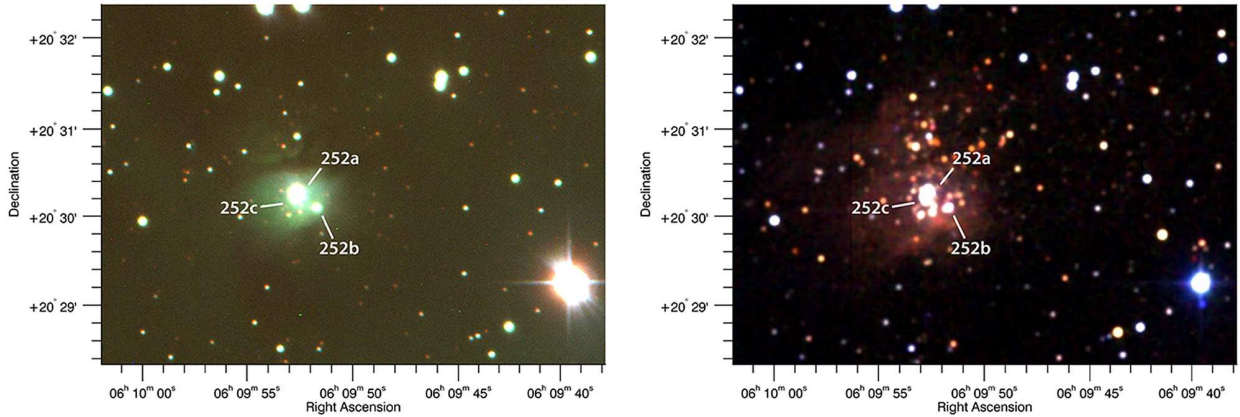}
\caption{\footnotesize Left: $VRI$ composite image from CTIO of the nebulosity and embedded cluster surrounding Sh~2-252a. This image was made from observations with the CTIO SMARTS 0.9 m telescope using the Johnson $V$ filter and the Kron-Cousins $R$ and $I$ filters, centered on 5438\AA, 6425\AA, and 8075\AA, respectively. DS9 software was used to create the color representation.  Note the dark dust bar that crosses the northern side of the nebulosity.
Right: $JHK_s$ composite image of the same field from 2MASS \citep{2masspaper}. North is up and east is to the left. J2000 coordinates are shown on both axes. Sh~2-252a, Sh~2-252b, and Sh~2-252c are identified. Sh~2-252a and Sh~2-252c are not resolved. The brightest star in the southwest corner is HD 42088.  \label{ctio}}
\end{figure}

\begin{figure}[htb!] 
\epsscale{0.85}
\centering
\plotone{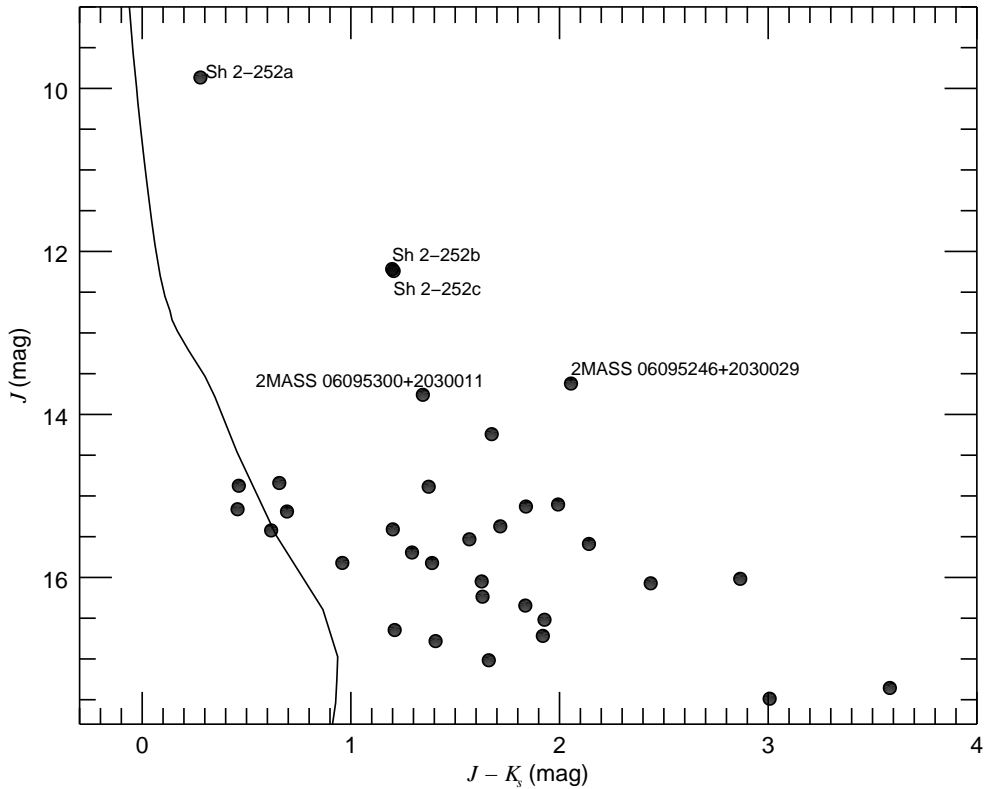}
\caption{\footnotesize Color-magnitude diagram of the sources within $30''$ of EPIC 202062176 using 2MASS magnitudes \citep{2masspaper}. The three stars targeted in this study are labeled, along with the next two brightest stars. The solid line represents the 5 Myr PARSEC main sequence isochrone of \citet{isochrone}, created using the distance ($d=1.4$ kpc), reddening ($E(B-V)=0.32$), and extinction relations of \citet{bonatto11}. \label{cmd} }
\end{figure}

We analyzed the K2 Campaign 0 light curve of EPIC~202062176 in order to determine the system's orbital elements.  However, there are over 100 main sequence and pre-main sequence stars in the embedded cluster surrounding EPIC~202062176 \citep{bonatto11}. The large K2 C0 photometric aperture actually records the flux of dozens of stars including several relatively bright and close by targets, seen in the $VRI$ and $JHK_s$ images in Figure~\ref{ctio}. Therefore, there is real uncertainty about which star in the vicinity of EPIC~202062176 is actually the eclipsing binary. In order to solve this problem, we obtained spectra of the three brightest stars within the K2 C0 aperture based on the color-magnitude diagram of these stars, shown in Figure \ref{cmd}. The names and coordinates of our targets are listed in Table~\ref{coord}.  

The brightest target in Sh~2-252E is Sh~2-252a, an early B-type star that corresponds to EPIC~202062176. Sh~2-252a has been studied in the past using both multi-band photometry and low resolution spectroscopy \citep{glz71, gc75, chavarria87, haikala95, jose12}. The second brightest target is Sh~2-252b, a Herbig Ae star located $16''$ southwest of Sh~2-252a \citep{chavarria89}. The third brightest target is a star $6''$ southeast of Sh~2-252a, which we call Sh~2-252c. The a and c components are not clearly resolved in Figure \ref{ctio}, but the pair is clearly visible in a higher resolution image presented in Figure~1 of the paper by \citet{chavarria87}.   

Our observations are described in Section \ref{observations}. We present our analysis of the K2 C0 light curve in Section \ref{photometry} and the spectroscopic analysis of Sh~2-252a, Sh~2-252b, and Sh~2-252c in Section \ref{spectroscopy}.  The results are summarized in Section \ref{discussion}.

\begin{deluxetable}{lcccl}
\tabletypesize{\scriptsize}
\tablewidth{\textwidth}
\tablecaption{Target Information\label{coord}}
\tablehead{ \colhead{Star}&  \colhead{$\alpha_{2000}$ }& \colhead{$\delta_{2000}$}& \colhead{$J$ (mag)} & \colhead{Other Identifications} }
\startdata
Sh~2-252a & 06:09:52.6 	& +20:30:16.4   & 9.86	
	&  2MASS 06095263+2030164\tablenotemark{a}, EPIC~202062176, ALS 17694\\	
Sh~2-252b & 06:09:51.7	& +20:30:06.2	& 12.22	
	&  2MASS 06095168+2030061, HBHA 2215-05  \\
Sh~2-252c & 06:09:52.8 	& +20:30:10.7  	& 12.24	
	&  2MASS 06095278+2030106	\\
\enddata
\tablenotetext{a}{\citet{lacourse15} list the incorrect 2MASS ID for EPIC 202062176 in their Table 2; the K2-TESS stellar properties catalogue \citep{k2tess} matches EPIC 202062176 with 2MASS 06095263+2030164, not 2MASS 06095262+2030273.}
\end{deluxetable}

\section{Observations}\label{observations}
\subsection{K2 Photometry}
\begin{figure}[htb!] 
\epsscale{0.9}
\centering
\plotone{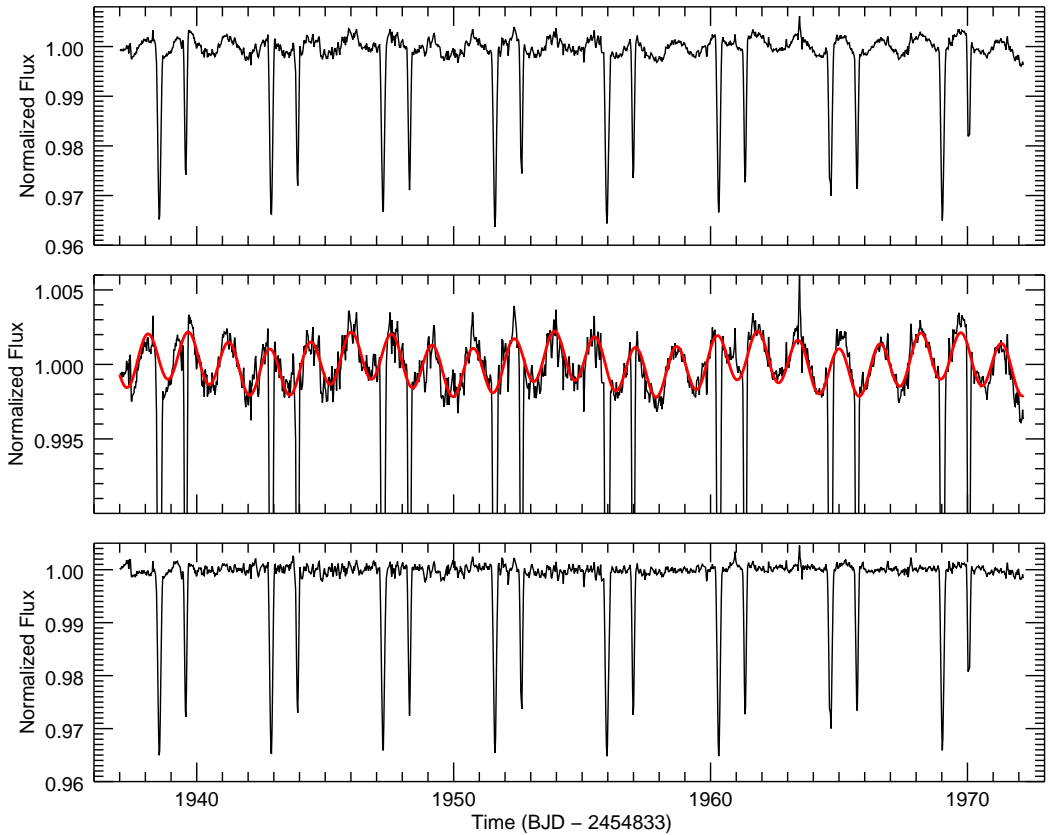}
\caption{\footnotesize Top: De-trended K2 C0 light curve of EPIC~202062176. Middle: Model of the non-orbital variations (red). Bottom: Variation-subtracted light curve. \label{pulsations} }
\end{figure}

EPIC~202062176 was observed in K2 Campaign 0 in long-cadence mode. We used the data from 2014 April 22 - 2014 May 27 (BJD 2456770 - 2456805), because data from the first half of Campaign 0 was lost due to a spacecraft pointing issue. We completed our data reduction using PyKE tools\footnote{http://keplerscience.arc.nasa.gov/software.html} \citep{pyke}. The light curve was extracted from a custom aperture mask, background subtracted, and de-trended. The aperture mask used was a circular mask roughly $40''$ in diameter and centered on EPIC 202062176. We corrected for spacecraft motion artifacts using the kepsff task of \citet{vanderburg}. The reduced light curve for EPIC~202062176 is shown in top panel of Figure \ref{pulsations}.

\subsection{Spectroscopy}
We obtained 13 nights of data for Sh~2-252a using the CHIRON echelle spectrograph \citep{chiron} on the CTIO 1.5 m telescope in 2014 November - December. We also obtained one night of data for Sh~2-252a, Sh~2-252b and Sh~2-252c using the Astrophysical Research Consortium echelle
spectrograph \citep[ARCES,][]{arces} on the APO 3.5 m telescope on 2016 February 11. The CHIRON spectrograph covers 4500-8900\AA \ at a resolving power of R$\sim$27400. Three 900 sec exposures were taken in fiber mode and averaged for each night. A thorium-argon lamp spectrum was taken after each object spectrum for wavelength calibration. These spectra were reduced using the CTIO pipeline from Yale University \citep{chiron}, transformed onto a heliocentric wavelength grid, and continuum normalized. The ARCES spectrograph covers 3500-10600\AA \ at a resolving power of R$\sim$31500. One spectrum was obtained for each target in slit mode. The exposure times for Sh~2-252a, Sh~2-252b, and Sh~2-252c were 1800 sec, 3200 sec, and 3800 sec, respectively. A thorium-argon lamp spectrum was taken after each object spectrum for wavelength calibration. These spectra were reduced using standard procedures in IRAF\footnote{IRAF is distributed by the National Optical Astronomy Observatory, which is operated by the Association of Universities for Research in Astronomy, Inc., under cooperative agreement with the National Science Foundation.}, transformed onto a heliocentric wavelength grid, and continuum normalized.

\section{Photometric Analysis}\label{photometry}

\subsection{Modeling the Non-orbital Variations}
\begin{figure}[htb!] 
\epsscale{0.9}
\centering
\plotone{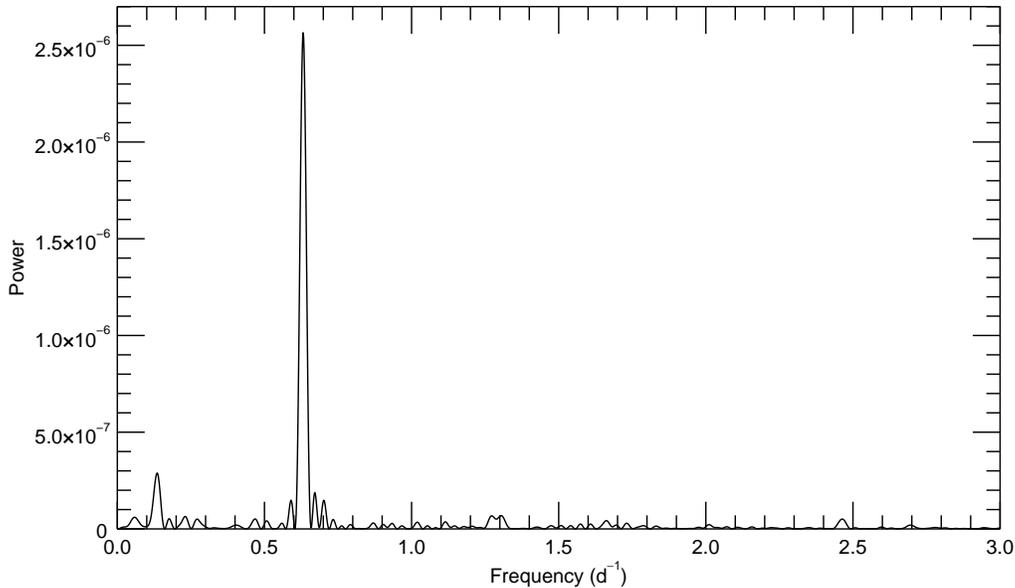}
\caption{\footnotesize Power spectrum of the non-orbital variations in the K2 C0 light curve. The two peaks correspond to frequencies $f_1 = 0.632$ d$^{-1}$ ($P_1 = 1.583$ d) and $f_2 = 0.134$ d$^{-1}$ ($P_2 = 7.477$ d).  \label{powerspec}}
\end{figure}

The K2 C0 light curve of EPIC~202062176 exhibits shallow eclipses as well as low amplitude variations, possibly caused by stellar pulsation or rotational modulation of star spots. We used the program Period04 \citep{period04} to determine these photometric periods by completing a Fourier transform of the eclipse-free light curve. The resulting power spectrum revealed only two strong periods, $P_1 = 1.583$ days and $P_2 = 7.477$ days,  as shown in Figure \ref{powerspec}. We subtracted a model of these periodicities from the observed light curve in order to isolate the eclipses, as shown in the middle panel of Figure \ref{pulsations}. 

\subsection{Modeling the Light Curve}
\begin{figure}[htb!]
\epsscale{0.9}
\centering
\plotone{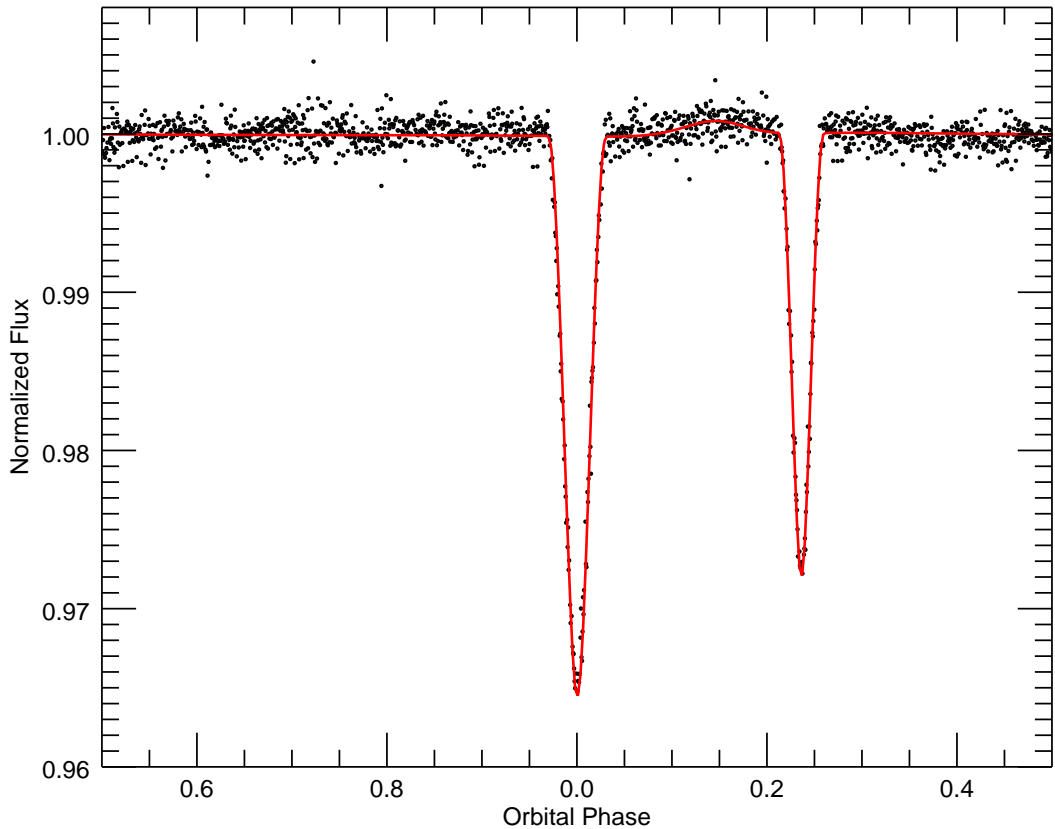}
\caption{ \footnotesize Phased K2 C0 light curve (black points) and best-fit ELC model (red, solid line). Phase 0 corresponds to the epoch of primary eclipse. \label{foldedlc} }  
\end{figure}

We folded the K2 C0 light curve based on the orbital period and epoch of primary eclipse found by \citet{lacourse15}, $P = 4.354$ days and $T_0 = $ BJD 2456775.899. Figure \ref{foldedlc} shows the folded light curve. From the eclipse durations and separations, we calculated the eccentricity, $e=0.45$, and longitude of periastron, $\omega=199^\circ$, which are consistent with the values found by \citet{lacourse15}.

We then used a light curve synthesis program, Eclipsing Light Curve code \citep[ELC,][]{ELC}, to model the eclipses and determine the orbital parameters of the eclipsing binary. The period, epoch of primary eclipse, eccentricity, and longitude of periastron were held fixed while we varied other parameters, such as the inclination, Roche lobe filling factors, and $T_{\rm eff}$ ratio. We also assumed a mass ratio $q = M_2/M_1 = 1$, although the solutions are relatively insensitive to this estimate. Table \ref{ELCtable} lists the best-fit parameter estimates, found by visual inspection. Figure \ref{foldedlc} shows the best-fit ELC model. 

The eclipse depths of the eclipsing binary are only 3-4\% percent, which implies a relatively low inclination. However, reducing the inclination was not enough to match the observed light curve. By fitting ELC models with different third light components, we found that a third light contribution of 90\% is needed to dilute accurately the eclipses, i.e., 90\% of the total flux comes from nebular and stellar light from Sh~2-252E while only 10\% comes from the binary itself.   However, there is a limited family of solutions due to the direct correlation between inclination and third light contamination: the inclination ranges from $i=84^\circ$ for a contamination factor of $82\%$, to $i=90^\circ$ for a contamination factor of $93\%$. 
The increase in flux between eclipses (at phase $=0.15$) due to tidal distortions provided a lower limit to the inclination and contamination factor, because the models with inclinations lower than $84^\circ$ and low contamination display a much larger increase in flux than that found in the observed light curve (Figure \ref{foldedlc}).
Future observations using higher resolution photometry should be able to reduce the amount of contamination and better determine the inclination of the binary system.

\begin{deluxetable}{ll}
\tablewidth{.69\textwidth}
\tablecaption{Preliminary Orbital Parameters of the Eclipsing Binary \label{ELCtable}}
\tablehead{ \colhead{Parameter} & \colhead{Estimated Value} }
\startdata
Orbital Period $P$ (days) \tablenotemark{a}				&\ \ 4.354250    \\
Epoch of primary eclipse $T_0$ (BJD) \tablenotemark{a}		&\ \ 2456775.8988    \\ 	
Epoch of periastron (BJD)								&\ \ 2456772.2194   \\ 	
Eccentricity $e$ 									&\ \ 0.45   \\
Longitude of periastron $\omega$ (deg)					&\ \ 199   \\
Inclination $i$ (deg)									&\ \ 86.5  \\
Effective temperature ratio $T_s/T_p$					&\ \ 0.90 \\
Relative radius of primary $r_p/a$						&\ \ 0.096    \\
Relative radius of secondary $r_s/a$ 					&\ \ 0.086  \\
Third light contamination	(\%)							&\ \ 90 \\
\enddata
\tablenotetext{a}{Fixed values from \citet{lacourse15}}
\end{deluxetable}

\clearpage

\section{Spectroscopic Analysis}\label{spectroscopy}

\subsection{Sh~2-252a}
\subsubsection{Stellar Properties}
\begin{figure}[htb!]
\epsscale{0.95}
\centering
\plotone{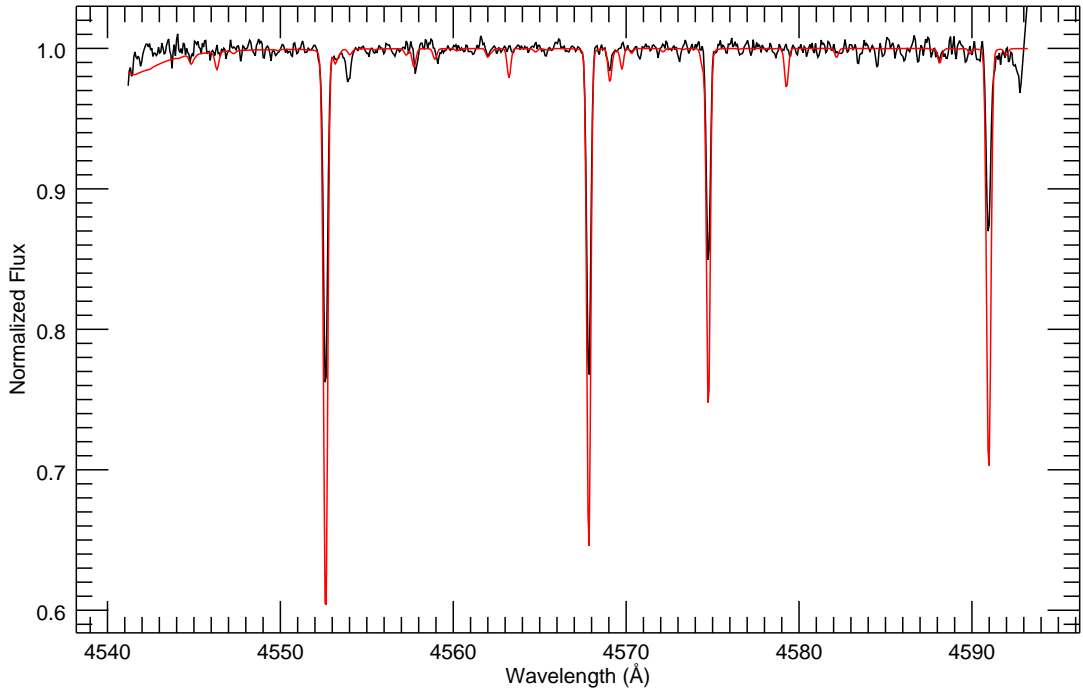}
\caption{ \footnotesize Time-averaged CHIRON spectrum (black) for Sh~2-252a and TLUSTY model spectrum (red) for $T_{\rm eff}$=29000 K, $\log g$=4.3, and $V \sin i=1$ km~s$^{-1}$, showing the dilution of the triplet \ion{Si}{3} $\lambda \lambda 4552, 4567, 4574$. \label{siIII}  } 
\end{figure}

Sh~2-252a was classified in the past between O8 - B1 V \citep{glz71, gc75, haikala95, jose12}. To confirm this spectral type, we compared our time-averaged CHIRON spectrum to the TLUSTY BSTAR2006 model spectra of \citet{tlustyb}. Because Sh~2-252a is only a few million years old \citep{bonatto11}, all models have solar abundances. Due to reflected continuum light from the surrounding nebula \citep{gc75}, all of the absorption lines are shallower in the observed spectrum than in the model spectra. To avoid problems of absolute line strength, we measured the ratios of the equivalent widths of several absorption lines in our observed spectrum and in TLUSTY models of various effective temperatures ($T_{\rm eff}$) and surface gravities ($\log g$).  We used the temperature-dependent ratios \ion{N}{3} $\lambda$4640/\ion{N}{2} $\lambda$4643, \ion{Si}{4} $\lambda$4654/\ion{Si}{3} $\lambda$4568, and \ion{C}{2} $\lambda$5696/\ion{Al}{3} $\lambda$5697 to determine the best-fit $T_{\rm eff}$, and used the luminosity-dependent ratios \ion{He}{1} $\lambda$4387/\ion{O}{2} $\lambda$4416, \ion{Si}{3} $\lambda$4552/\ion{He}{1} $\lambda$4387, and \ion{Si}{4} $\lambda$4089/\ion{He}{1} $\lambda$4026 to determine the best-fit $\log g$. We found that Sh~2-252a has $T_{\rm eff} = 29000\pm900$ K and $\log g = 4.3 \pm 0.1$, where the errors correspond to the standard deviation of the results from all line ratios. Therefore, Sh~2-252a corresponds to a spectral type of B0.5 V, according to Table B.1 in \citet{gray}, which is consistent with past studies. Figure \ref{siIII} shows the observed triplet \ion{Si}{3} $\lambda \lambda 4552, 4567, 4574$ as well as a TLUSTY model for $T_{\rm eff} = 29000$ K. We then used this model spectrum to measure the third light contamination of the CHIRON spectra to be 34\% and of the ARCES spectra to be 12\%. Because the width of the ARCES slit ($1\farcs6$) is smaller than the diameter of the CHIRON fiber ($2\farcs7$), much less reflected continuum light was able to enter the spectroscopic aperture of ARCES. 

Figure \ref{siIII} also demonstrates how sharp are the absorption lines of Sh~2-252a. We found the projected rotational velocity ($V \sin i$) of Sh~2-252a by first scaling the line depths of the model spectrum to match the depths of the CHIRON spectrum. We then measured the instrumental broadening of the ThAr lamp spectra to be 12 km~s$^{-1}$ FWHM and fit the observed line profiles over models of different $V \sin i$. We found that $V \sin i<10$ km~s$^{-1}$, i.e., any projected rotational velocity is too small to measure at the resolving power of our spectra.

The spectra of Sh~2-252a show strong nebular emission lines, such as the Balmer series and several forbidden lines, from the diffuse H II region (NGC 2174) and nebulosity (Sh~2-252E) near Sh~2-252a. \citet{gc75} found that the Balmer lines originate from both NGC 2174 and Sh~2-252E, while the [\ion{N}{2}] and [\ion{O}{3}] lines originate only from Sh~2-252E. We measured the equivalent width ($W_\lambda$) of H$\alpha$ emission in the spectra of Sh~2-252a for comparison with that of Sh~2-252b and Sh~2-252c, listed in Table \ref{RVtable}. H$\alpha$ was saturated in one or more of the three exposures for each night of CHIRON observations, so we only used exposures where H$\alpha$ is unsaturated. Nights when all three exposures are saturated are left blank in the table. This measurement was made using numerical integration across the emission line profile, with the baseline determined by a linear fit between endpoints of the line. Sh~2-252a is much brighter than Sh~2-252b or Sh~2-252c, so the nebular emission lines do not appear as strong relative to the bright continuum of Sh~2-252a. 

\subsubsection{Radial Velocity}
We measured the radial velocity ($V_r$) of each CHIRON spectrum using cross-correlation with a TLUSTY model spectrum of $T_{\rm eff}=29000$ K. We combined the results of multiple echelle orders to calculate the average $V_r$ for each night, listed in Table \ref{RVtable}. The average radial velocity from all 13 nights is 25.109 $\pm$ 0.025 km~s$^{-1}$. Sh~2-252a does not show significant radial velocity variations on the orbital timescale of the eclipsing binary, and therefore must not be part of the binary system. The radial velocity of the ARCES spectrum is  $V_r = 25.18 \pm 0.24$ km~s$^{-1}$. Even though this spectrum was taken 14 months after the CHIRON spectra, the radial velocities are very consistent. Sh~2-252a could be a distant companion to the binary system, but would have an orbital period much longer than our 14 month baseline.

\begin{deluxetable}{ccccrc}
\tablewidth{0.9\textwidth}
\tablecaption{Radial Velocity Measurements \label{RVtable}}
\tablehead{ \colhead{Star} & \colhead{Date} & \colhead{$V_r$} & \colhead{$\sigma_{V_r}$} & \colhead{$-W_\lambda(H\alpha)$ } & \colhead{Observation} \\  
\colhead{} & \colhead{(HJD)}  & \colhead{(km~s$^{-1}$)} & \colhead{(km~s$^{-1}$)} & \colhead{(\AA) } & \colhead{Source}}
\startdata					
Sh~2-252a	 &	2456980.604	  &	24.91	 &	0.09	 &	2.54	 	&	CTIO \\
Sh~2-252a	 &	2456982.656	  &	25.04	 &	0.08	 &	2.40	 	&	CTIO \\
Sh~2-252a	 &	2456986.882	  &	25.22	 &	0.09	 &	2.32	 	&	CTIO \\
Sh~2-252a	 &	2456987.707	  &	25.14	 &	0.09	 &	2.49	 	&	CTIO \\
Sh~2-252a	 &	2456988.704	  &	25.14	 &	0.10	 &	\nodata	 &	CTIO \\
Sh~2-252a	 &	2456990.571	  &	25.14	 &	0.10	 &	\nodata	 &	CTIO \\
Sh~2-252a	 &	2456991.612	  &	25.11	 &	0.12	 &	\nodata	 &	CTIO \\
Sh~2-252a	 &	2456992.871	  &	25.13	 &	0.09	 &	\nodata	 &	CTIO \\
Sh~2-252a	 &	2456994.604	  &	25.07	 &	0.10	 &	\nodata	 &	CTIO \\
Sh~2-252a	 &	2456995.895	  &	25.04	 &	0.08	 &	2.25	 	&	CTIO \\
Sh~2-252a	 &	2456997.680	  &	24.99	 &	0.09	 &	2.69	 	&	CTIO \\
Sh~2-252a	 &	2457012.859	  &	25.09	 &	0.11	 &	\nodata	 &	CTIO \\
Sh~2-252a	 &	2457013.771	  &	25.29	 &	0.10	 &	2.90	 	&	CTIO \\
Sh~2-252a	 &	2457430.670 	  &	25.18	 &	0.24	 &	2.81		&	APO	 \\ \\
Sh~2-252b 	 &	2457430.700	  &	16.80	 &     2.99	 &	157.56	&	APO \\ \\
Sh~2-252c	 &	2457430.744	  &	17.44	 &	1.28	 &	80.19	&	APO	\\
\enddata
\end{deluxetable}

To investigate whether the eclipsing binary is another star still within the spectroscopic aperture, we searched for hidden spectral features of the binary system using a Doppler tomography algorithm \citep{tomography}. We first calculated the semi-major axis of the binary system from Kepler's Third Law, based on the orbital period found by \citet{lacourse15} and assuming both components are $1  M_\odot$ stars ($q=M_2/M_1 = 1$). Combining this with the inclination, eccentricity, and time of periastron from ELC, we calculated the velocity semi-amplitude of the binary system. Finally, we used the SBCM program \citep{sbcm} to predict the radial velocity of each component at the time of each CHIRON observation, using the same systemic velocity as that of Sh~2-252a. The Doppler tomography algorithm then worked to find any observed spectral lines with these radial velocity shifts and created a reconstructed spectrum with these lines. Unfortunately, our reconstructed spectrum of the binary star did not show any spectral features, so we conclude that the eclipsing binary is more separated from Sh~2-252a than the fiber aperture radius, i.e., more than $1\farcs4$ away.

\clearpage

\subsection{Sh~2-252b}
\subsubsection{Stellar Properties}
\begin{figure}[ht!]   
\epsscale{1}
\centering
\plotone{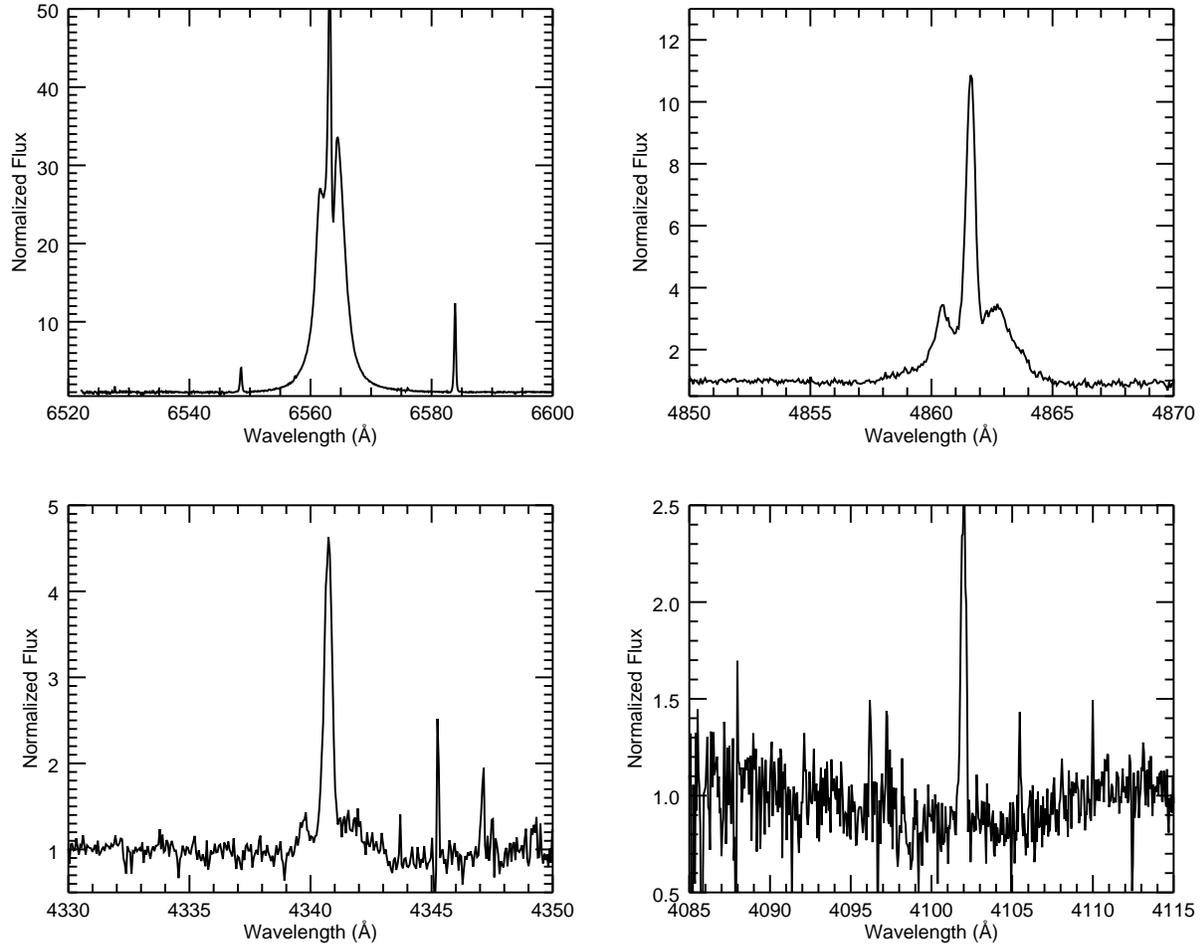}
\caption{ \footnotesize Balmer lines of Sh~2-252b. Top left: H$\alpha$ $\lambda$6563. Top right: H$\beta$ $\lambda$4861. Bottom left: H$\gamma$ $\lambda$4340. Bottom right: H$\delta$ $\lambda$4100. The strong, narrow center peaks originate in the diffuse H II region (NGC 2174). The broad profiles of H$\alpha$ and H$\beta$ are characteristic of a Herbig Ae/Be star, as identified by \citet{chavarria89}. H$\delta$ shows evidence of broad photospheric absorption from Sh~2-252b. \label{balmer}  } 
\end{figure}

\citet{chavarria89} discovered that Sh~2-252b is a Herbig Ae star, based on its broad H$\alpha$ emission, infrared flux excess, and high luminosity. Our ARCES spectrum of Sh~2-252b shows broad Balmer and Paschen emission lines, as well as narrow nebular emission from NGC 2174 and Sh~2-252E. H$\alpha$ has a double-peaked profile, shown in the top left panel of Figure \ref{balmer}, indicative of a circumstellar disk. We measured the equivalent width of H$\alpha$ in the same manner as described above and found $W_\lambda$ = -158\AA. Note that this equivalent width does include nebular emission components.

Sh~2-252b does not show any strong absorption lines,  likely due to dilution from bright circumstellar material \citep{gray-corbally}, making spectral classification very difficult. \citet{chavarria89} created an SED from broad-band photometry and measured the Balmer jump to determine the spectral type of Sh~2-252b to be A0 $\pm$ 2 subclasses. We instead estimated the Johnson $B-V$, $V-R$, and $V-I$ colors for Sh~2-252b based on the intrinsic colors of a B0.5 V star from \citet{wegner94} and the raw flux values ($F$) of Sh~2-252a and Sh~2-252b at 4400\AA, 5500\AA, 6410\AA, and 7890\AA.  For example, the equation for the $B-V$ color of Sh~2-252b is given by 
\begin{displaymath}
(B-V)_{252b} = (B-V)_{252a} - 2.5 \log \frac{(F_{252b}/F_{252a})_B}{(F_{252b}/F_{252a})_V} \\
\end{displaymath}
where $(B-V)_{252a} = -0.24$ \citep{wegner94}. This approach assumes that the observations were made with approximately the same air mass and seeing conditions. We estimated the colors of Sh~2-252b to be $B-V=-0.02$, $V-R=0.01$, and $V-I= 0.02$, which all match those of an A0 V star \citep{wegner94}.

\subsubsection{Radial Velocity}
\begin{figure}[h!] 
\centering
\epsscale{0.9}
\plotone{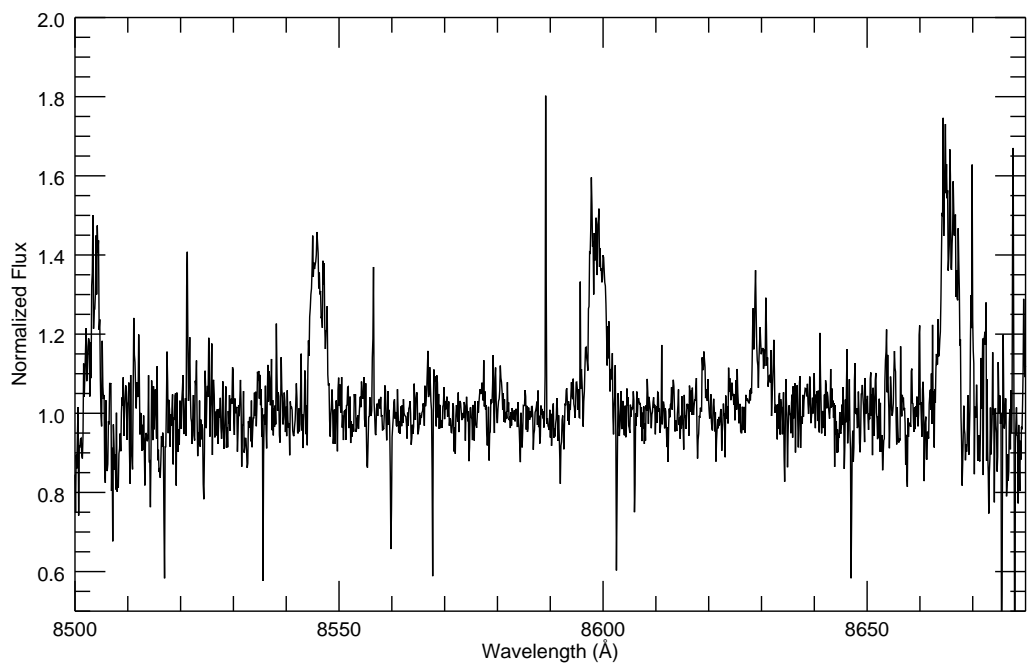}
\caption{ \footnotesize Example Paschen lines of Sh~2-252b; from left to right, P16 $\lambda$8502, P15 $\lambda$8545, P14 $\lambda$8598,  and P13 $\lambda$8665. These were among the emission lines used to calculate the radial velocity of Sh~2-252b. \label{paschen} } 
\end{figure}

Because Sh~2-252b does not show absorption lines, we estimated its radial velocity using the Paschen emission lines in the near-IR, shown in Figure \ref{paschen}. We fit a single Gaussian to each emission line, and found that the average $V_r$ is 16.8 km~s$^{-1}$ with a standard deviation of 3.0 km~s$^{-1}$. We then used the SBCM code to predict the radial velocities of the eclipsing binary stars at the time of our ARCES observation. The predicted radial velocities of two $1 M_\odot$ stars in the binary system with the same systemic velocity of Sh~2-252a are $+63$ and $-13$ km~s$^{-1}$, which does not match the radial velocity of Sh~2-252b.  We conclude that Sh~2-252b is not a component of the eclipsing binary system.

\clearpage

\subsection{Sh~2-252c}
\subsubsection{Stellar Properties}
Sh~2-252c has only been observed in the past as part of the 2MASS survey \citep{2masspaper}, but is relatively faint and has poor quality $JHK_s$ magnitudes. No prior spectroscopic observations or spectral classifications have been made for this star. Unfortunately, Sh~2-252c does not show absorption lines in the blue due to the low signal-to-noise ratio of our spectrum, so we were unable to identify the \ion{Ca}{2} H \& K lines, the G band, or \ion{Ca}{1} $\lambda4226$ line which could all aid in spectral classification \citep{gray-corbally}. The red and near-IR proved more useful. First, there is no evidence of broad hydrogen absorption under the nebular emission, which suggests that Sh~2-252c is a cooler, lower mass star. Second, we were able to identify several \ion{Fe}{1}, \ion{Ca}{1}, and \ion{Ca}{2} absorption lines, as well as possible \ion{Fe}{2}, \ion{Ti}{1}, and \ion{Li}{1} absorption lines. Specifically, \ion{Fe}{1} $\lambda \lambda 5328, 6137, 6400, 8220, 8691$, \ion{Fe}{2} $\lambda 8228$, \ion{Li}{1} $\lambda 6708$, \ion{Ti}{1} $\lambda \lambda 8382, 8435$, \ion{Ca}{1} $\lambda \lambda 6102, 6122, 6162, 6439, 6450, 6462$, and the triplet \ion{Ca}{2} $\lambda \lambda 8498, 8544, 8664$. These absorption lines are likely diluted by reflected continuum light from Sh~2-252E, as was the case for Sh~2-252a. However, many of these absorption lines are present in F through M stars, so we could only conclude that Sh~2-252c is a mid-F or later type star. 

We also calculated the $B-V$, $V-R$, and $V-I$ colors of Sh~2-252c in the same manner as for Sh~2-252b. We found $B-V=0.41$, corresponding to a mid-F star, $V-R=0.48$, corresponding to a early K star, and $V-I= 1.1$, corresponding to an early K star \citep{pecaut13}. This discrepancy in spectral types implies that Sh~2-252c has an infrared excess or blue deficient flux. Furthermore, we used the raw flux values to estimate $V \approx 15$ for Sh~2-252c. Comparing the $V$ magnitudes of Sh~2-252a and Sh~2-252c to their $J$ magnitudes from 2MASS, Sh~2-252c is overly bright in $J$ by almost 2 magnitudes. This suggests that Sh~2-252c is a pre-main sequence star, which is consistent with the appearance of the \ion{Ca}{1} and \ion{Ca}{2} absorption lines. These lines are very narrow compared to the same lines in the Sun's spectrum, due to the decreased pressure broadening of a pre-main sequence star relative to a main sequence dwarf. 

\subsubsection{Radial Velocity}
\begin{figure}[htb!]
\epsscale{1}
\centering
\plotone{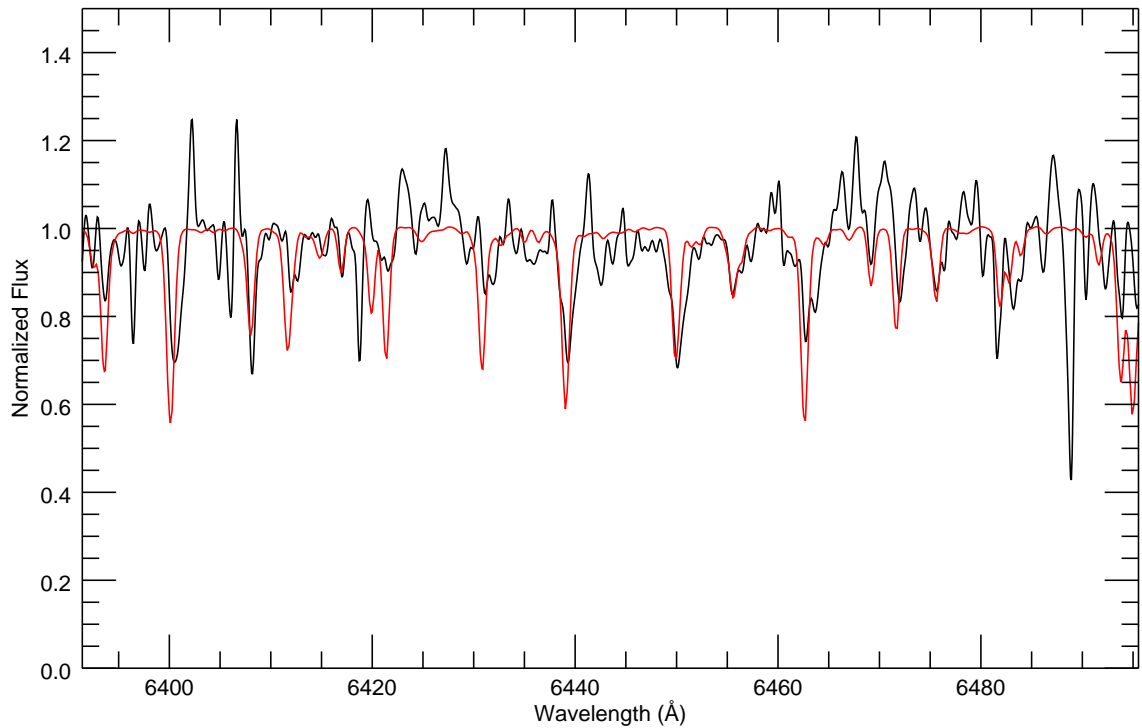}
\caption{ \footnotesize Smoothed spectrum of Sh~2-252c (black), compared to a BLUERED model (red) of $T_{\rm eff}=5000$ K, $\log g$=4.5, and solar metallicity. This echelle order was one of three used for cross-correlation to determine the radial velocity of Sh~2-252c. \label{order72}  } 
\end{figure}

To calculate the radial velocity of Sh~2-252c, we cross-correlated our ARCES spectrum with a BLUERED model spectrum from \citet{bluered}.  We used a model with $T_{\rm eff}=5000$ K, corresponding to a K2 star and the approximate spectral type of Sh~2-252c. An example part of the spectrum is shown in Figure \ref{order72}, which shows only marginal agreement with the observed spectrum due to the low signal to noise ratio of our spectrum and contamination with emission components in some lines. We found $V_r =17.4 \pm 1.3$ km~s$^{-1}$ for Sh~2-252c. As described above, the predicted $V_r$ of the eclipsing binary stars at the time of our ARCES observation are $+63$ and $-13$ km~s$^{-1}$. We do not find any absorption lines with these radial velocities nor do we see any evidence of double lines, which excludes Sh~2-252c as a possible component of the eclipsing binary system.

\section{Discussion}\label{discussion}
We conclude that EPIC~202062176 is not actually the eclipsing binary itself, but its measured flux is contaminated by a nearby binary in the K2 C0 aperture. The eclipses are highly diluted by light from nearby stars and the surrounding nebulosity within the large aperture, with the binary contributing only 10\% of the total flux. We also found that the component stars are similar in effective temperature and therefore are likely similar masses. This young, eccentric binary would be an interesting target for studies on apsidal motion due to its young, low mass components and close orbit. Assuming $q=1$, the semi-major axis of the eclipsing binary system is $a= 14.1 R_\odot ((M_p+M_s)/(2 M_\odot))^{1/3}$. Combining this with the relative radii determined by ELC, the individual stellar radii are thus $r_p = 1.4 R_\odot$ and $r_s = 1.2 R_\odot$. The primary and secondary stars are larger than the expected main sequence radius and are likely pre-main sequence stars. The periastron distance of the two stars is $7.8 R_\odot$, much larger than $r_p + r_s = 2.6R_\odot$, suggesting that the two components are completely detached. 

Further work to identify the eclipsing binary could include time series photometry with better spatial resolution than Kepler or spectroscopic observations of the other nearby stars. We ruled out the three brightest stars as possible members of the eclipsing binary, so other candidates are the next two brightest stars within the photometric aperture: 2MASS 06095246+2030029 ($J=13.6$) and 2MASS 06095300+2030011 ($J=13.8$).  These stars are labeled on the color-magnitude diagram in Figure \ref{cmd}. The remaining nearby stars are very faint ($J>14$) and do not contribute enough light to be probable members of the eclipsing binary system. 

The other intriguing element of the K2 C0 light curve was the low amplitude variation, with $P_1 = 1.583$ days and $P_2 = 7.477$ days. One explanation is that these variations are caused by pulsations of Sh~2-252a. The pulsation periods are too long for the $p$-mode pulsations of $\beta$ Cephei and $\delta$ Scuti stars, but the periods are consistent with the $g$-mode pulsations of $\gamma$ Doradus and slowly-pulsating B stars \citep{astroseismology}. However, the spectral type of Sh~2-252a is much earlier than the usual spectral types of these pulsating stars. A more plausible explanation would be that the observed variations are caused by the rotational modulation of star spots. Sh~2-252a is not a likely source, because its slow rotation would create variations with much longer periods. On the other hand, these periods are reasonable for the rotation of Sh~2-252b and Sh~2-252c.

Even though we excluded Sh~2-252a (EPIC~202062176), Sh~2-252b, and Sh~2-252c as components of the eclipsing binary system, these stars are still very interesting. Sh~2-252a is a B0.5~V star with remarkably sharp absorption lines. We found $V \sin i <10$ km~s$^{-1}$, which is very slow even for a low inclination. This is very surprising for such a young, early type star that would have been born with high angular momentum from the molecular cloud.
 It is possible that the low $V\sin i$ is due to a small inclination, but the probability is low that we view the star with a pole-on orientation. Alternatively, Sh~2-252a could have been spun down by a magnetic field, the case for many early-type slow rotators. For example, \citet{petit13} found that three of the four hottest B stars with magnetic fields are very sharp-lined. One of these is $\tau$ Sco, a well studied magnetic B0.2~V star with $V \sin i < 13$ km~s$^{-1}$ \citep{tau sco}, similar to Sh~2-252a. Furthermore, about 10\% of B stars have magnetic fields \citep{bfield}, and young massive stars are even more likely to have magnetic fields \citep{bdecay}. Therefore, Sh~2-252a is  possibly a magnetic star due to its slow rotation and young age.  If it is magnetic, the strong magnetic field required to slow its rotation in only 5 Myr would likely be detectable, making Sh~2-252a a promising target for the study of magnetic O and B stars. The magnetic field of Sh~2-252a could also have influenced the accretion disk during its formation and even confined the outflow of debris to create the dark dust band seen in Figure \ref{ctio}. 
Next, Sh~2-252b is a Herbig A0e star with a double-peaked H$\alpha$ profile, evidence of a circumstellar disk. Sh~2-252c is likely a low mass pre-main sequence star due to its IR excess, over-brightness, and possible dilution from circumstellar gas and dust. While we concluded that these stars are not components of the binary system itself, we cannot exclude the possibility that they could be a distant third companion to the binary system. Nonetheless the eclipsing binary in Sh~2-252E was clearly born in a dense environment with remarkable neighbors, and further efforts to identify the binary among the cluster stars will help elucidate its role in the early dynamical evolution of the cluster.

\acknowledgments
{
We would like to thank Todd Henry for obtaining the CTIO images and Suzanne Hawley for providing us with discretionary observing time at APO. Kepler was competitively selected as the tenth Discovery mission. Funding for this mission is provided by NASA's Science Mission Directorate. K2 data were obtained from the Mikulski Archive for Space Telescopes (MAST). STScI is operated by the Association of Universities for Research in Astronomy, Inc., under NASA contract NAS5-26555. Support for MAST for non-HST data is provided by the NASA Office of Space Science via grant NNX09AF08G and by other grants and contracts.    This work also made use of PyKE (Still \& Barclay 2012), a software package for the reduction and analysis of Kepler data. This open source software project is developed and distributed by the NASA Kepler Guest Observer Office.    This material is based upon work supported by the National Science Foundation under Grant No.~AST-1411654. Institutional support has been provided from the GSU College of Arts and Sciences and from the Research Program Enhancement fund of the Board of Regents of the University System of Georgia, administered through the GSU Office of the Vice President for Research and Economic Development.}

{\it Facilities:} \facility{Kepler/K2}, \facility{ARC}, \facility{CTIO}

\appendix



\begin{thebibliography}{}

\bibitem[Aerts et al.(2010)]{astroseismology} Aerts, C., Christensen-Dalsgaard, J., \& Kurtz, D.~W.\ 2010, Asteroseismology (Dordrecht: Springer)

\bibitem[Armstrong et al.(2015)]{armstrong15} Armstrong, D.~J., Kirk, J., Lam, K.~W.~F., et al.\ 2015, \aap, 579, A19 

\bibitem[Bagnuolo et al.(1992)]{tomography} Bagnuolo, W.~G., Jr., Gies, D.~R., \& Wiggs, M.~S.\ 1992, \apj, 385, 708 

\bibitem[Bertone et al.(2008)]{bluered} Bertone, E., Buzzoni, A., Ch{\'a}vez, M., \& Rodr{\'{\i}}guez-Merino, L.~H.\ 2008, \aap, 485, 823  

\bibitem[Bonatto \& Bica(2011)]{bonatto11} Bonatto, C., \& Bica, E.\ 2011, \mnras, 414, 3769 

\bibitem[Chavarria-K.~et al.(1987)]{chavarria87} Chavarria-K., C., de Lara, E., \& Hasse, I.\ 1987, \aap, 171, 216 

\bibitem[Chavarria-K.~et al.(1989)]{chavarria89} Chavarria-K., C., Leitherer, C., de Lara, E., Sanchez, O., \& Zickgraf, F.-J.\ 1989, \aap, 215, 51 

\bibitem[Chen et al.(2015)]{isochrone} Chen, Y., Bressan, A., Girardi, L., et al.\ 2015, \mnras, 452, 1068 

\bibitem[Conroy et al.(2014)]{conroy14} Conroy, K.~E., Pr{\v s}a, A., Stassun, K.~G., et al.\ 2014, \pasp, 126, 914 

\bibitem[Felli et al.(1977)]{felli77} Felli, M., Habing, H.~J., \& Israel, F.~P.\ 1977, \aap, 59, 43 

\bibitem[Fossati et al.(2016)]{bdecay} Fossati, L., Schneider, F.~R.~N., Castro, N., et al.\ 2016, \aap, 592, A84 

\bibitem[Garnier \& Lortet-Zuckermann(1971)]{glz71} Garnier, R., \& Lortet-Zuckermann, M.~C.\ 1971, \aap, 14, 408 

\bibitem[Grasdalen \& Carrasco(1975)]{gc75} Grasdalen, G.~L., \& Carrasco, L.\ 1975, \aap, 43, 259 

\bibitem[Gray(2008)]{gray} Gray, D.~F.\ 2008, The Observation and Analysis of Stellar Photospheres, 3rd ed. (Cambridge, UK: Cambridge University Press)

\bibitem[Gray \& Corbally(2009)]{gray-corbally} Gray, R.~O., \& Corbally, C. J.\ 2009, Stellar Spectral Classification (Princeton, NJ: Princeton University Press)

\bibitem[Grunhut et al.(2012)]{bfield} Grunhut, J.~H., Wade, G.~A., \& MiMeS Collaboration \ 2012, in Stellar Polarimetry: From Birth to Death, AIP Conf. Vol. 1429, ed. J. L. Hoffman, J. Bjorkman, \& B. Whitney (Melville, NY: AIP), 67

\bibitem[Haikala(1995)]{haikala95} Haikala, L.~K.\ 1995, \aap, 294, 89  

\bibitem[Howell et al.(2014)]{howell14} Howell, S.~B., Sobeck, C., Haas, M., et al.\ 2014, \pasp, 126, 398 

\bibitem[Jose et al.(2012)]{jose12} Jose, J., Pandey, A.~K., Ogura, K., et al.\ 2012, \mnras, 424, 2486 

\bibitem[LaCourse et al.(2015)]{lacourse15} LaCourse, D.~M., Jek, K.~J., Jacobs, T.~L., et al.\ 2015, \mnras, 452, 3561 

\bibitem[Lanz \& Hubeny(2007)]{tlustyb} Lanz, T., \& Hubeny, I.\ 2007, \apjs, 169, 83

\bibitem[Lenz \& Breger(2005)]{period04} Lenz, P., \& Breger, M.\ 2005, Communications in Asteroseismology, 146, 53  

\bibitem[Morbey \& Brosterhus(1974)]{sbcm} Morbey, C.~L., \& Brosterhus, E.~B.\ 1974, \pasp, 86, 455 

\bibitem[Orosz \& Hauschildt(2000)]{ELC} Orosz, J.~A., \& Hauschildt, P.~H.\ 2000, \aap, 364, 265 

\bibitem[Pecaut \& Mamajek(2013)]{pecaut13} Pecaut, M.~J., \& Mamajek, E.~E.\ 2013, \apjs, 208, 9 

\bibitem[Petit et al.(2013)]{petit13} Petit, V., Owocki, S.~P., Wade, G.~A., et al.\ 2013, \mnras, 429, 398 

\bibitem[Sharpless(1959)]{sharpless59} Sharpless, S.\ 1959, \apjs, 4, 257 

\bibitem[Sim{\'o}n-D{\'{\i}}az et al.(2006)]{tau sco} Sim{\'o}n-D{\'{\i}}az, S., Herrero, A., Esteban, C., \& Najarro, F.\ 2006, \aap, 448, 351 

\bibitem[Skrutskie et al.(2006)]{2masspaper} Skrutskie, M.~F., Cutri, R.~M., Stiening, R., et al.\ 2006, \aj, 131, 1163 

\bibitem[Stassun et al.(2014)]{k2tess} Stassun, K.~G., Pepper, J.~A., Oelkers, R., et al.\ 2014, arXiv:1410.6379 

\bibitem[Still \& Barclay(2012)]{pyke} Still, M., \& Barclay, T.\ 2012, PyKE: Reduction and analysis of Kepler Simple Aperture Photometry data, Astrophysics Source Code Library, ascl:1208.004 

\bibitem[Tokovinin et al.(2013)]{chiron} Tokovinin, A., Fischer, D.~A., Bonati, M., et al.\ 2013, \pasp, 125, 1336 

\bibitem[Wang et al.(2003)]{arces} Wang, S.-i., Hildebrand, R.~H., Hobbs, L.~M., et al.\ 2003, \procspie, 4841, 1145 

\bibitem[Wegner(1994)]{wegner94} Wegner, W.\ 1994, \mnras, 270, 229 

\bibitem[Vanderburg \& Johnson(2014)]{vanderburg} Vanderburg, A., \& Johnson, J.~A.\ 2014, \pasp, 126, 948 

\end{thebibliography}
\end{document}